\pdfoutput=1
\documentclass[preprint,aps,showpacs,preprintnumbers,amsmath,amssymb]{revtex4}
\usepackage{graphicx}
\usepackage{dcolumn}
\usepackage{bm}
\usepackage{color}
\begin{document}
\title{Phosphorene confined systems in magnetic field, quantum transport, 
and superradiance in the
quasi-flat band}
\author{B. Ostahie$^{1,2}$, and A. Aldea$^{1}$}
\affiliation{$^1$ National Institute of Materials Physics, 
77125 Bucharest-Magurele, 
Romania \\
$^2$ Faculty  of Physics, University of Bucharest, Romania}
\date{\today}

\begin{abstract}
Spectral and transport properties of 
electrons in confined phosphorene 
systems are investigated in a five hopping parameter tight-binding model, 
using analytical and numerical techniques.  The  main emphasis is on 
the properties of the topological edge states  accommodated by the quasi-flat band
that characterizes the  phosphorene energy spectrum.

We show, in the particular case of  phosphorene,  how the breaking
of the bipartite lattice structure gives rise to the electron-hole asymmetry
of the energy spectrum. The properties of  the  topological
edge states in the zig-zag nanoribbons 
are analyzed under different aspects: degeneracy, localization,
extension in the Brillouin zone, dispersion of the quasi-flat band 
in magnetic field. The finite-size phosphorene plaquette exhibits 
a Hofstadter-type spectrum made up of two unequal butterflies 
separated by a gap, where a quasi-flat band composed of  zig-zag edge 
states is located.
The transport properties are investigated by simulating a four-lead 
Hall device (importantly, all leads are attached on the same zig-zag side),
and using the Landauer-B\"uttiker formalism. We find out that 
the chiral edge states due to the  magnetic field yield  quantum Hall 
plateaus, but  the topological edge states in the gap do not support 
the quantum Hall effect and prove a dissipative behavior.  By calculating the 
complex eigenenergies of the non-Hermitian effective Hamiltonian that 
describes the open system (plaquette+leads), we prove the superradiance 
effect in the energy range of the quasi-flat band, with consequences 
for the density of states and electron transmission properties.
\end{abstract}
\pacs{73.20.At, 73.63.-b, 73.43.-f }
\maketitle
\section{Introduction}
The very recent revival of the black phosporene physics
comes from the  technical possibility to obtain monolayers,
known as phosphorene, with specific topological properties.
Phosphorene is a quasi-2D structure organized
as a  puckered hexagonal lattice,  the top and side views being 
shown in Fig.1a and Fig.1b, respectively.
One may think that, due to the structural similarity, the
electron properties of phosphorene are   resembling
those of graphene. However, in contradistinction to graphene,
the phosphorene is an  anisotropic direct gap semiconductor,
much more attractive for electronic devices.
Beside the monolayered structure, multilayers of black phosporous are
also studied, mainly in order to control
the band gap, in the perspective of a  potential application
for  field-effect transistors.

The phosphorene ribbon geometry (especially, with zig-zag edges)
 is also conceptually
interesting since, instead of the
semi-metallic spectrum of graphene, distinguished by a flat band at $E=0$,
and extending  between the points K and K' in the Brillouin zone (BZ),
the phosphorene shows well-separated valence and conduction bands
and a {\it quasi}-flat band in the middle of the gap,  composed of
edge states that exists at any momentum $k\in BZ$.

In the tight-binding model, the phosphorene lattice is described by
five hopping integrals $t_1,t_2,..,t_5$ \cite{Katsnelson}, which induce
the  significant differences in the electron spectrum that are 
noticed when
compared  to graphene. The model points out also the anisotropy
of the energy spectrum: both the top  of the valence band and the
bottom of the conduction band look
quadratically as function of $k_y$ , but
nearly linear as function of $k_x$ (Dirac-like) (see Fig.2), situation which
is described in terms of hybrid Dirac spectrum \cite{Montambaux, Goerbig}.
The hopping integral $t_4$ plays a distinctive role as it connects sites
of the same kind on the hexagonal lattice, breaking the bipartitism of the  
lattice, and, as a consequence, the electron-hole symmetry of the energy
spectrum is also broken \cite{Mielke}.
As an additional effect due to  $t_4$,  we shall see in Sec.II that
the  edge states, organized in a perfect flat band at $t_4=0$, undergoes
dispersion in the case of nonvanishing $t_4$.

The properties of the macroscopically degenerate
flat (quasi-flat) bands composed of edge states
in confined systems (ribbon or finite-size plaquette) attract much 
attention nowadays, and phosphorene presents a serious advantage 
coming from  the existence of a gap that protects the quasi-flat band,
such that its properties can be  evidenced in a cleaner way.
The study of the spectral and transport properties in the magnetic field,
and the response of the quasi-flat band to
the invasive contacts of a Hall device, identified as a superradiant 
phenomenon, are a topic of our paper.

Similar to graphene, the confined phosphorene exhibits two
types of edge states: i) the {\it chiral} edge states generated 
by a strong perpendicular magnetic field $\mathcal{B}$, and
supporting the quantum Hall effect (QHE), and ii) the edge states 
typical to the zig-zag boundaries  in the hexagonal lattice, 
which exists even  in the absence of the magnetic field. 
The last ones, which will be called {\it topological} edge states \cite{Ezawa}, 
are non-chiral and remain like that even at $\mathcal{B} \neq 0$. 
Obviously, they do not show QHE, but show longitudinal conductance, 
i.e., they have a dissipative character.

The transport calculations  assume the knowledge of the full 
Hamiltonian of the open system consisting of the finite-size system 
of interest (namely, the phosphorene plaquette) and the semi-infinite 
leads. Technically speaking, one uses actually an {\it effective} 
Hamiltonian obtained by formal elimination of the degree of freedom 
of the leads, however, as the price to be paid, the result is a 
non-Hermitian Hamiltonian with complex eigenvalues. 
The method of non-Hermitian Hamiltonian has been used for the 
calculation of transport properties of the quantum dots in
the Landauer-B\"{u}ttiker formalism (see for
instance \cite{Moldoveanu}), but also in the localization-delocalization
problem in the 1D non-Hermitian Anderson model \cite{Nelson,Khoruz,Celardo}.
In these two different problems, the non-hermicity arises from 
different sources, however we do not enter here such  peculiar aspects.

In the phosphorene confined system, the complex eigenvalues of the
effective Hamiltonian in the energy range of the quasi-flat band, 
corroborated by the calculation of the electron transmission, 
density of states and local density of states make evident  
a specific superradiant behavior of the topological edge states.
(We remind that the superradiance consists in the segregation of 
eigenenergies and  overlapping of some resonances, the process 
being controlled by the lead-system coupling \cite{Nesterov}.)
For instance, the density of states of the  quasi-flat band
exhibits a miniband structure,
each miniband behaving as a 1D-conducting channel
with the conductance $G=e^2/h$. These aspects are discussed in Sec.IV.

Some quantum transport aspects in phosphorene were very recently revealed.
The quantum Hall effect and spin splitting of the Landau levels (LL) were
observed in \cite {KaiChang, Likai},
and also Shubnikov-deHaas oscillations of the longitudinal
resistance were found in \cite{Likai1}.
The transport anisotropy, reflecting  the structural one, was shown
experimentally by measuring the angle dependence of the drain
current \cite{HanLiu} or by the non-local response \cite{Levitov}.
The strain-induced modifications of the phosphorene band structure were studied
in \cite{Neto,Fey}.
The field effect transistor is also the topic of \cite{Wu,Likai2}.

The paper is organized as follows. Sec.II presents the
tight-binding Hamiltonian, Peierls phases in magnetic field and discusses the
question of electron-hole symmetry breaking. Sec.III is devoted to the
study of the phosphorene ribbon in the magnetic field, as an extension of Ezawa
analysis at $\mathcal{B}=0$. Sec.IV deals with the spectral and transport
properties of the phosphorene mesoscopic plaquette, showing the specific
features of the quantum Hall effect in the bands, and the properties resulting
from the  superradiance effect in the quasi-flat band.
The summary and conclusions can be found in the last section. 

\section{The tight-binding model and  electron-hole symmetry breaking in 
phosphorene} 
Similar to graphene, the unit cell contains two atoms called $A$ and $B$, 
however the phosphorene  tight-binding Hamiltonian is more complicated 
as it contains five hopping integrals to nearest and
next-nearest neighbors. In order to write down the Hamiltonian 
we define the creation and annihilation operators $a^\dag_{nm},a_{nm},
b^\dag_{nm},a_{nm}$, where $n$ and $m$ are cell indexes along the Ox- and
Oy-axis, respectively.
In the presence of a perpendicular magnetic field, which will be  
described by the vector potential 
$\vec{\mathcal{A}}=(-\mathcal{B}y,0,0)$, 
some of the hopping integrals acquire the Peierls phase  expressed 
by the circulation of the vector potential along the trajectory 
connecting the two end points :
\begin{equation}
\phi_{AB}=\frac{2\pi }{\Phi_0}\int_A^B\vec{\mathcal{A}}d \vec{l}
=-\frac{2\pi  B}{\Phi_0}\int_{x_A}^{x_B} y(x) dx.
\end{equation}
Much attention should be paid to the calculation of the phases since
the angle $\beta$ describing the deviation from the perfect flat 
2D lattice (see Fig.1b) enter also the calculation.
Since  the hopping integral $t_4$ plays the  special role 
mentioned in the introduction, we separate the terms proportional 
to this parameter, and the {\it spinless} tight-binding Hamiltonian  
of the phosphorene lattice under perpendicular magnetic field
will be written as follows:
\begin{eqnarray}
H&=&H^0+H^4 ,
\nonumber \\
H^0&=&\sum_{nm}E_a a^\dag_{nm}a_{nm}+E_b b^\dag_{nm}b_{nm}+ 
t_1 \big(e^{i\phi_1}a^\dag_{n+1m}+e^{-i\phi_1}a^\dag_{nm}\big)b_{nm}
+t_2a^\dag_{nm+1}b_{nm} 
\nonumber \\
&+&t_3\big(e^{i\phi_3}a^\dag_{nm+2}+e^{-i\phi_3}a^\dag_{n-1m+2})b_{nm}  
+t_5 a^\dag_{n+1m-1}b_{nm} + H.c.~,
\nonumber \\
H^4&=&\sum_{nm}t_4 \big(e^{i\phi_{4B}}b^\dag_{nm+1}+
e^{-i\phi_{4B}}b^\dag_{n-1m+1}\big)b_{nm}
\nonumber \\
&+&t_4 \big(e^{i\phi_{4A}}a^\dag_{nm+1}+e^{-i\phi_{4A}}a^\dag_{n-1m+1}\big)a_{nm}
+H.c.,
\end{eqnarray}
where $E_a$ and $E_b$ are the atomic energies at the sites $A$ and $B$, 
respectively, and , according to \cite{Katsnelson}, 
$t_1=-1.220 eV , t_2=3,665 eV, t_3=-0.205 eV, t_4=-0.105 eV, t_5=-0.055 eV$.
Since the hopping parameters are given in electron-volts, all
the quantities in the paper having the dimension of energy will be measured 
in the same units.

\begin{figure}[!ht]
\centering
\includegraphics[scale=1.0]{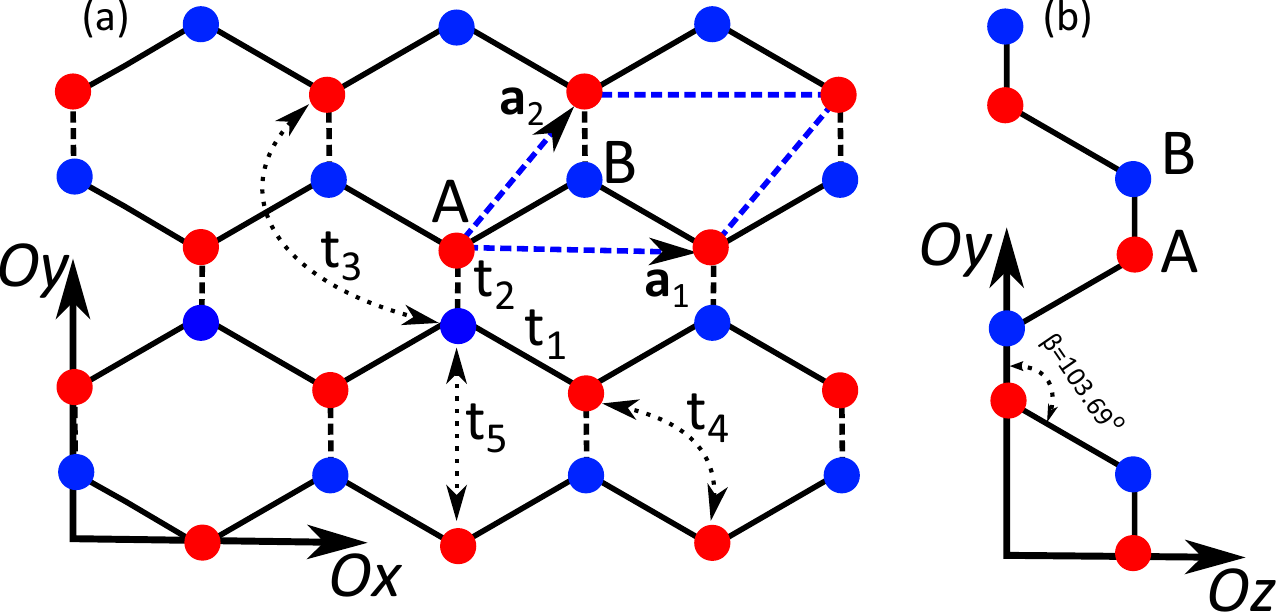}
\caption{(Color online) (a) 
Schematic representation of phosphorene lattice  with 
two types of edges,  zig-zag (along the  $\textbf{O}_{x}$ direction) 
and armchair (along the $\textbf{O}_{y}$ direction); 
$t_1, t_2, t_3, t_4, t_5$ are the hopping amplitudes that connect 
the lattice sites;  $\textbf{A}$ (red) and $\textbf{B}$ (blue) index 
the two types of atoms, and the dashed blue lines represent the unit cell 
with $\textbf{a}_1$ and $\textbf{a}_2$ as unit vectors. 
(b) The projection of the lattice on the yz-plane. The number of lattice sites 
is $7\times4$.}
\end{figure}

In the chosen gauge of the  vector potential, only three hopping
integrals acquire a Peierls phase in  magnetic field, 
namely $t_1,t_3$ and $t_4$. For instance, $\phi_1(m)$ in Eq.(1)
is the Peierls phase corresponding to the hopping from the site
B in the cell (n,m) to the site A in the next cell (n+1,m), and
equals:
\begin{equation}
\phi_1(m)=2\pi  \frac{\mathcal{B}}{\Phi_0} 
\int_{B_{nm}}^{A_{n+1,m}} y(x) dx=
-2\pi \frac{\Phi}{\Phi_0}\frac{1}{6}\big((m-1)(1+2 sin\beta)-\frac{1}{2}
\big). \nonumber
\end{equation}
Similarly, the other phases in Eq.(2) are \cite{Note-phase},
\begin{eqnarray}
\phi_3(m)&=& -2\pi \frac{\Phi}{\Phi_0}\frac{1}{6}\big( m(1+2 sin\beta)-
\frac{1}{2} \big)
\nonumber\\
\phi_{4A}(m)&=& -2\pi \frac{\Phi}{\Phi_0}\frac{1}{6}( m-\frac{1}{2})
(1+2 sin\beta)
\nonumber\\
\phi_{4B}(m)&=& -2\pi \frac{\Phi}{\Phi_0}\frac{1}{6}\big( (m-\frac{1}{2})
(1+2 sin\beta)+1\big),
\nonumber
\end{eqnarray}
where $\Phi/\Phi_0$ is the magnetic flux through the hexagonal cell
measured in quantum flux units, and $\beta$ is the angle shown in Fig.1b.
One notices that the phases $\phi_{4A}$ and $\phi_{4B}$ acquired by $t_4$ 
along  the A-A and B-B link, respectively, are different.

The spectral properties of the Hamiltonian (2)  can be studied
under different boundary conditions describing different geometries 
as the infinite sheet, the ribbon or the finite plaquette.
The  phoshorene infinite sheet can be simulated assuming  periodic 
boundary conditions along the both directions $Ox$ and $Oy$.
Let us consider first the case $\mathcal{B}=0$,
and use the Fourier transform of the 
creation and annihilation operators:
\begin{eqnarray}
a_{nm}=\sum_{\vec{k}}a_{\vec{k}}e^{i\vec{k}\vec{R}_{nm}},~~
b_{nm}=\sum_{\vec{k}}b_{\vec{k}}e^{i\vec{k}\vec{R}_{nm}},~~
\vec{R}_{nm}=n\vec{a}_1+m \vec{a}_2,
\end{eqnarray}
which helps in writing the Hamiltonian as a $2\times 2$ matrix in the 
momentum space $\vec{k}=(k_x,k_y)$:
\begin{equation}
H=\sum_{\vec{k}}H^0_{\vec{k}}+H^4_{\vec{k}}=\sum_{\vec{k}}
    \left(\begin{array}{cc}
      a^{\dagger}_{\vec{k}}~ & b^{\dagger}_{\vec{k}} 
\end{array}\right)
        \left(\begin{array}{ccc}
                T^4(\vec{k}) & T^0(\vec{k}) \\
                T^{0*}(\vec{k}) & T^4(\vec{k})
        \end{array}\right)
    \left(\begin{array}{c}
            a_{\vec{k}} \\
            b_{\vec{k}}  \\
\end{array}\right),
\end{equation}
with 
\begin{eqnarray}
T^0(\vec{k})&=&t_1(1+e^{-i\vec{k}\vec{a}_1})+t_2e^{-i\vec{k}\vec{a}_2}
+t_3(e^{-i2\vec{k}\vec{a_2}}+e^{i\vec{k}\vec{a_1}-i2\vec{k}\vec{a_2}})
+t_5 e^{-i\vec{k}\vec{a_1}+i\vec{k}\vec{a_2}}
\nonumber\\
T^4(\vec{k})&=&2t_4(cos\vec{k}\vec{a_2}+cos\vec{k}(\vec{a_1}-\vec{a_2})).
\end{eqnarray}

In approaching the question of spectrum symmetries, we remind that the
{\it electron-hole symmetry} of an energy spectrum holds if there 
exists an operator $\mathcal{P}$ that anticommutes with the Hamiltonian, 
$\{H,\mathcal{P}\}_+=0$. 
Indeed, it is quite straightforward to see that, if $E$ is an eigenvalue, 
$H\Psi_E= E\Psi_E$, then the energy  $-E$ belongs also to the spectrum, 
the corresponding eigenfunction being 
$\tilde\Psi_{-E}=\mathcal{P}\Psi_E$.
For our specific  problem of phosphorene, let us consider the operator
\begin{equation}
\mathcal{P}=\sum_{\vec{k}} a^\dag_{\vec{k}}a_{\vec{k}}- 
b^\dag_{\vec{k}}b_{\vec{k}}~.
\end{equation}
Obviously, this operator anticommutes with $H^0$, attesting that the 
energy spectrum of $H^0$ is electron-hole symmetric, 
$\{E_k^0,-E_k^0\}\in Sp$.
However, $\mathcal{P}$ does not anticommute with the total 
Hamiltonian $H^0+H^4$, the result being proportional to $t_4$.
One concludes that the phosphorene spectrum is electron-hole symmetric
if $t_4=0$, i.e, when one forgets about the hopping to the next-nearest 
neighbors, but it is not necessarily symmetric otherwise. 

The energy spectrum of the Hamiltonian (4)
can  be obtained analytically from the characteristic equation 
\begin{equation}
\begin{vmatrix}
E-T^4(\vec{k})  &  T^0(\vec{k})\\
T^{0*}(\vec{k}) & E-T^4(\vec{k})\\ 
\end{vmatrix}
=0 ,
\end{equation}
resulting a two-band spectrum of semiconducting type, with the
eigenvalues 
\begin{equation}
E_{\pm}(\vec{k})=T^4(\vec{k})\pm |T^0(\vec{k})| .
\end{equation}
The above equation confirms that for  $t_4=0$  the  spectrum becomes
symmetric $E=\pm |T^0(\vec{k})|$, with  a direct gap  
at the $\Gamma$ point equal to
$\Delta=2|T^0(0)|=2(2t_1+t_2+2t_3+t_5)$, 
which is  Ezawa's result \cite{Ezawa}.
On the other hand, Eq.(8)   proves that a nonvanishing $t_4$ shifts 
the whole  spectrum with $4t_4$, such that the electron-hole symmetry 
around $E=0$ is lost.
One concludes that the spectral asymmetry in phosphorene is 
the consequence of the hopping parameter $t_4$, which  connects 
sites of the same type and violates in this way the bipartitism of 
the lattice.
\begin{figure}
\includegraphics[scale=0.5]{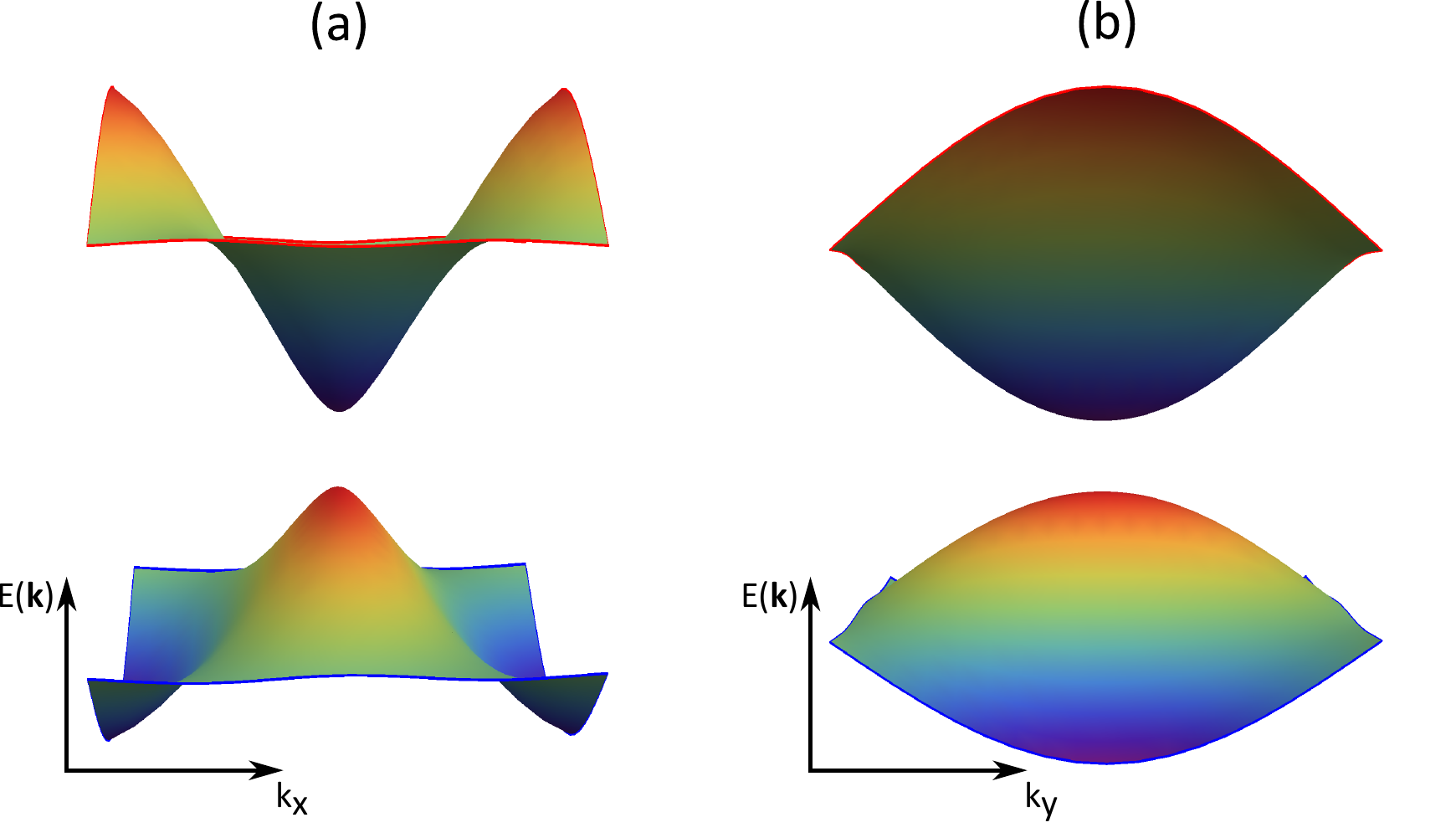}
\caption{(Color online)  The energy spectrum of the phosphorene lattice 
with periodic boundary conditions.  The anisotropy of the spectrum around 
the $\Gamma$ point  can be  observed: (a) the energy dispersion  
along the $k_x$-axis shows the Dirac-like behavior, and (b)
energy dispersion along the $k_y$-axis shows the Schr\"odinger-like behavior.} 
\end{figure}
The eigenvalues Eq.(8) are displayed in Fig.2,  where three aspects
have to be noticed: the presence of the gap, the strong anisotropy, and
the electron-hole asymmetry of the bands. As explained in \cite{Montambaux},
the first two properties occur in the hexagonal-type lattice as soon as 
$t_1\neq t_2$, even neglecting the other hopping parameters
in the Hamiltonian. 

The Hofstadter spectrum generated by a perpendicular magnetic field 
is also  very specific, consisting of two unequal butterflies 
separated by a gap.
There is no agreement yet on the field dependence of the 
Landau levels, as in \cite{KaiChang, Pereira} the dependence is linear, 
while in \cite{Ezawa1} the dependence is  $\sim~\mathcal{B}^{2/3}$.
In Fig.3 we show the numerically calculated Hofstadter spectrum  of
a {\it finite (mesoscopic) } plaquette, which exhibits a {\it supplementary} 
band in the middle accommodating the edge states \cite{KaiChang1}. 
The narrow width of the band, and the weak dependence on the magnetic field 
should be noticed. 
One has to observe that the spectrum misses the known periodicity with the
magnetic flux $E(\Phi+\Phi_0)=E(\Phi)$, which is met in the case of 
the 2DEG in perpendicular magnetic field. This comes from the presence 
of three different Peierls phases in the Hamiltonian (2).
The case of the phosphorene finite plaquette will be discussed  in Sec.III.

The Hofstadter-type spectrum in Fig.3 suggests new physical 
properties of the edge states, and stimulates a more extensive study of 
the phosphorene mesoscopic systems. They are simulated in 
the tight-binding model by imposing vanishing boundary conditions for 
the wave function all along the perimeter (the case of the finite-size 
plaquette) or only along two parallel zig-zag edges (the ribbon case).
\begin{figure}[!ht]
\centering
\includegraphics[angle=-90,scale=0.4]{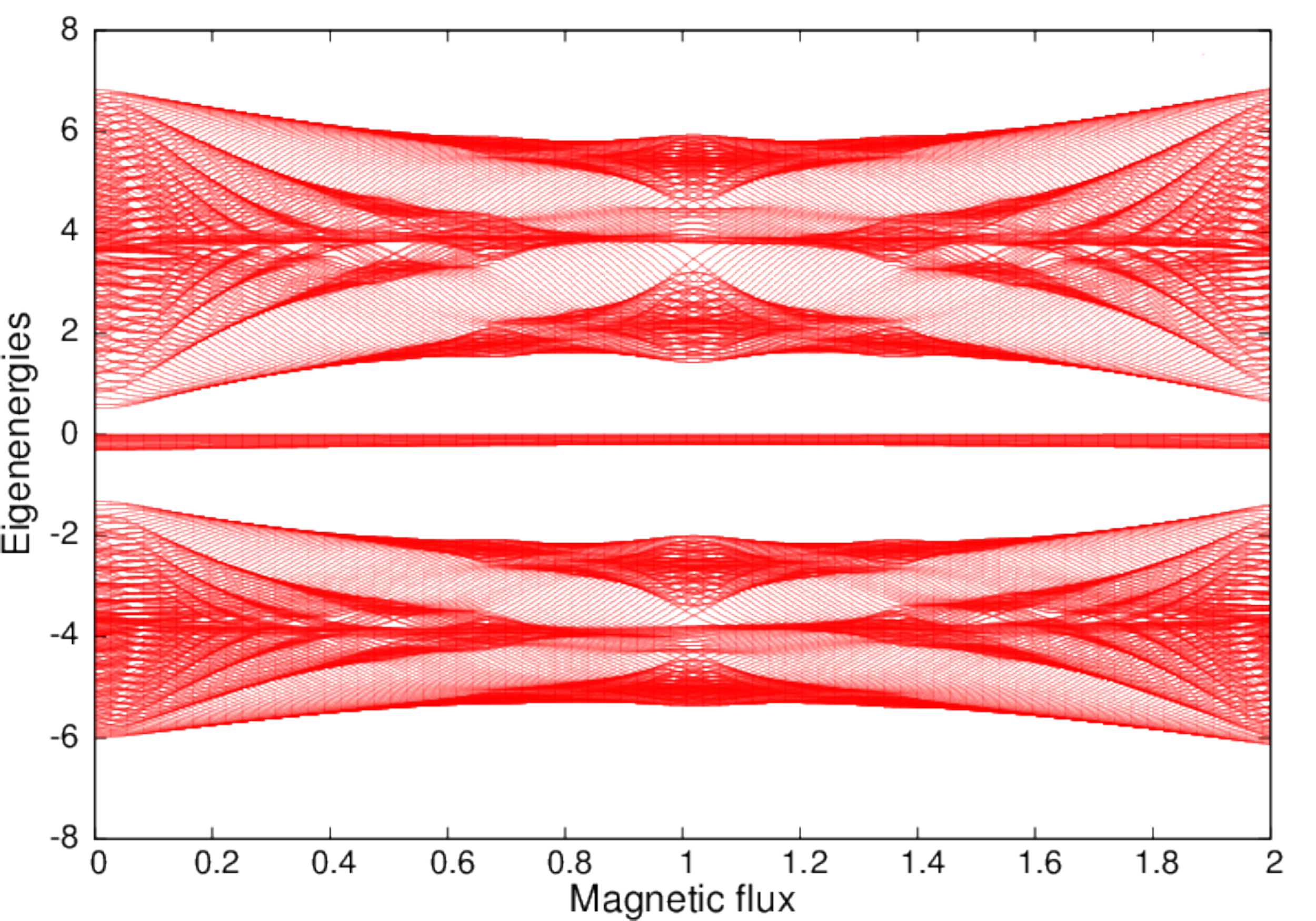}
\caption{(Color online) The Hofstadter spectrum of the finite 
phosphorene lattice. The quasi-flat band which accommodates topological
edge states can be noticed in the gap.
The number of lattice sites is $21\times20$, and the magnetic 
flux is measured in flux quanta $h/e$. }
\end{figure}

The phosphorene ribbon in the absence of the magnetic field is discussed in
\cite{Ezawa,Fazileh, Carvalho}, where it is shown that the zig-zag edges induce 
eigenvalues in the middle of the gap.
The band is perfectly flat (i.e., independent of the momentum $k$)
if $t_4=0$, and get  a slight dispersion 
otherwise. At a given $k$, there are two quasi-degenerate states 
which become perfectly degenerate in the limit of wide ribbons 
(similar to the case of graphene). 

The next section is devoted to the spectral properties of phosphorene  
ribbon in the presence of the magnetic field, in which case some 
new aspects  of interest can be proved even analytically.

\section{Spectral properties of the phosphorene ribbon in 
magnetic field}
Let us consider the Hamiltonian (2) and impose two edges parallel 
to the zig-zag chains at m=1 and m=M,  but keeping periodic 
boundary conditions along the x-direction. The Fourier transform 
along the x-direction gives rise to the following Hamiltonian for 
the ribbon geometry (where $k$ stands for $k_x$):
\begin{eqnarray}
\hskip-1cm
H&=&\sum_k H^0(k)+ H^4(k),
\nonumber \\
H^0(k)&=&\sum_{m=1}^M E_a a^\dag_{km}a_{km}+E_b b^\dag_{km}b_{km}
+ t_1 \big(e^{i(\phi_1-k)}+ e^{-i\phi_1} \big)a^\dag_{km}b_{km}
\nonumber \\
&+&t_2\sum_{m=1}^{M-1} a^\dag_{km+1}b_{km}
+t_3\sum_{m=1}^{M-2}\big(e^{i\phi_3}+
e^{-i(\phi_3-k)}\big)a^\dag_{km+2}b_{km} 
+t_5\sum_{m=1}^M  e^{-ik} a^\dag_{km-1}b_{km} + H.c.~~
\nonumber\\
H^4(k)&=&t_4\sum_{m=1}^{M-1} \big(e^{i\phi_{4B}}+
e^{-i(\phi_{4B}-k)}\big)b^\dag_{km+1}b_{km}
+ \big(e^{i\phi_{4A}}+
e^{-i(\phi_{4A}-k)}\big)a^\dag_{km+1}a_{km}  + H.c.~.
\end{eqnarray}

In the case of vanishing magnetic field $\mathcal{B}=0$, the 
energy spectrum of the above Hamiltonian  
is described by Ezawa \cite{Ezawa}. 
The numerical calculation takes into account all the five hopping 
integrals, but the analytical one considers 
$t_3=t_5=0$, while the parameter $t_4$ is considered perturbatively. 
The existence of a quasi-flat band in the gap, whose dispersion  
comes from $t_4$, is proved (see Eq.(22)in \cite{Ezawa}). 
We reobtain this result, which is shown in Fig.5(left), in order to be compared 
with the case of nonvanishing magnetic field in Fig.5.(right)

\subsection{The quasi-flat band in magnetic field}
The aim of this subsection is to elucidate the effect of the 
perpendicular magnetic field on the  spectral properties of the
edge states in the ribbon geometry. The formation of the quasi-flat 
band in the middle of 
the gap and the interesting degeneracy lifting due to the magnetic field 
are put forward both numerically and analytically.

Let us consider the atomic energies $E_a=E_b=0$, and the hopping
integrals $t_3=t_5=0$ (as in Ref.5), but keep $\mathcal{B }\ne 0$.
Then, $H^0(k)$ becomes:
\begin{equation}
H^0(k)=\sum_{m=1}^M 
 t_1 \big(e^{i(\phi_1-k)}+ e^{-i\phi_1}  \big)a^\dag_{km}b_{km}
+\sum_{m=1}^{M-1}t_2 a^\dag_{km+1}b_{km}  + H.c. ~.
\end{equation}
For any momentum $k$, we look for the eigenfunctions of $H^0(k)$ as
\begin{equation}
|\Psi^0(k)\rangle=\sum_{m=1}^M(\xi^A_{km}  a^\dag_{km}+
\xi^B_{km} b^\dag_{km})~|0\rangle,
\end{equation}
and, from  $H^0 |\Psi^0(k)\rangle=E^0(k)|\Psi^0(k)\rangle$,
the equations satisfied by the coefficients  $\xi^{A,B}_{km}$ 
can be identified easily as: 
\begin{eqnarray}
t_1(e^{i\phi_1}+e^{-i(\phi_1-k)})\xi^A_{km}+
t_2\xi^A_{km+1}&=&E^0(k)\xi^B_{km}
\nonumber\\
t_1(e^{-i\phi_1}+e^{i(\phi_1-k)})\xi^B_{km}+
t_2\xi^B_{km-1}&=&E^0(k)\xi^A_{km}~, 
\end{eqnarray}
with $m=1,..,M$, and the ribbon-type  boundary conditions 
$\xi^B_{k,0}=0, \xi^A_{k,M+1}=0$.

We approach the study of the edge states in a way similar to the 
graphene case \cite{Wakabayashi}, i.e. assume the 
existence of a perfectly flat
band in the middle of the spectrum $E^0(k)=0$, and examine the 
properties of $\xi^A_{km}$ and $\xi^B_{km}$.
With the notation $\overline{t}_1(m)=t_1(e^{-i\phi_1}+e^{i(\phi_1-k)})$, 
Eqs.(12) provide
\begin{eqnarray}
\xi^A_{k,m}&=&\xi^A_{k,1}(-\overline{t}_1^*/t_2)^{m-1}, \nonumber\\
\xi^B_{k,m}&=&\xi^B_{k,M}(-\overline{t}_1/t_2)^{M-m}, 
\end{eqnarray}
where $\xi^A_{k,1}$ and $\xi^B_{k,M}$ can be obtained from 
the normalization condition.
Since $|\overline{t}_1/t_2|< 1$, it is obvious that $\xi^A_{k,m}$ 
reaches its maximum value at the edge $m=1$ and the minimum at the 
other edge $m=M$, while $\xi^B_{k,m}$ behaves oppositely.
This means that
$|\Psi^0_A(k)\rangle =\sum_m \xi^A_{km}a^\dag_{km}|0\rangle$
describes an edge state localized near the 
edge $m=1$, while
$|\Psi^0_B(k)\rangle =\sum_m \xi^B_{km}b^\dag_{km}|0\rangle$ 
is localized
at the other edge $m=M$; the two functions are obviously orthogonal. 
It is important to underline that,  for a {\it finite} width ribbon,
\{$|\Psi_A\rangle,|\Psi_B\rangle$\} are only {\it approximate} 
eigenfunctions of $H^0$ (corresponding to the {\it approximate} eigenvalue  
$E^0(k)=0$ ); this is evident from the fact that the matrix element 
$\langle\Psi_A|H^0(k)| \Psi_B\rangle\neq 0$ at any finite $M$.
Indeed, using Eq.(13), a straightforward calculation yields:
\begin{equation}
\langle\Psi_A|H^0(k)|\Psi_B\rangle= -t_2(-\overline{t}_1/t_2)^M~ \xi^{*A}_{k1}
\xi^B_{kM}~ , ~~~~ k\in(0,2\pi].
\end{equation}
The above result indicates that the  actual eigenfunctions 
describing  the edge states in the finite ribbon  consist of a 
superposition of the functions $|\Psi_A\rangle$ and $|\Psi_B\rangle$,  
the  eigenvalues being $E^0_{\pm}(k)=\pm 
t_2|\overline{t}_1/t_2|^M ~|\xi_{k1}^{A*}\xi_{kM}^B|$. Taking again 
into account  the convergence condition $|\overline{t_1}/t_2|<1$  
(which in the case of phosphoene is ensured at any $k$), one notices 
that the splitting $\delta_k =E^0_+(k)-E^0_-(k)$ 
vanishes  exponentially in the limit $M\rightarrow\infty$. 
Only in this limit $E^0(k)=0$ becomes a double-degenerate 
non-dispersive (flat) band, similar to the situation in graphene, 
but with the notable contradistinction that, for phosphporene,  
this property is true for any momentum $k$ \cite{NoteW}.
We checked also numerically the energy splitting $\delta$ as 
function of the ribbon width, and the exponential decay with 
increasing $M$ is obvious in Fig.4. 
\begin{figure}[!ht]
\centering
\includegraphics[scale=0.6]{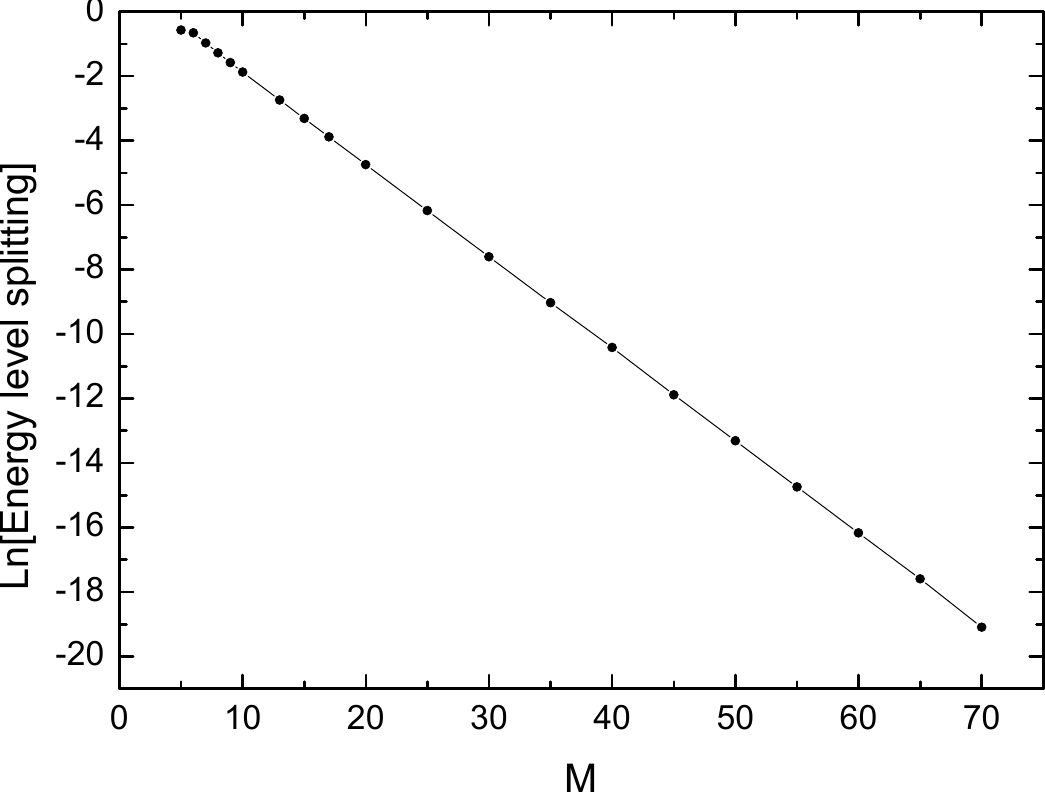}
\caption{The numerically calculated energy splitting, 
at the $\Gamma$ point $k=0$ and vanishing magnetic field $\mathcal{B}=0$,
as function of the ribbon width $M$.  The calculation takes into account all the
five hopping parameters in the Hamiltonian (9).}
\end{figure}

\begin{figure}[!ht]
\centering
\includegraphics[scale=0.7]{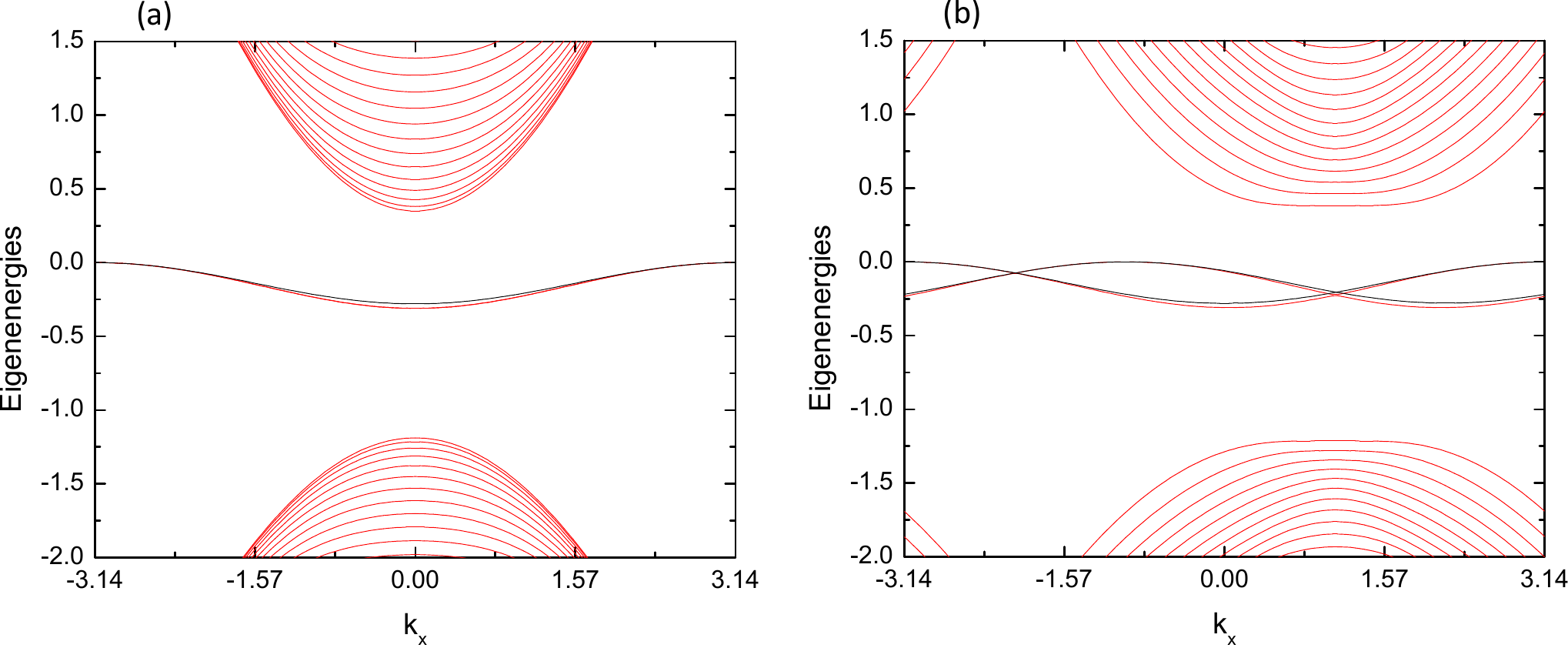}
\caption{(Color online)
The low energy spectrum of the phosphorene zig-zag nanoribbon. 
The degeneracy lifting of the quasi-flat band induced by the magnetic field
can be noticed by comparing the  two panels : in the left panel  the magnetic 
flux is zero, while in the right panel  $\Phi=0.01\Phi_0$. 
The red lines represent the numerically calculated spectrum for the width $M=71$, 
the black lines represent the analytical results Eq. (17) and Eq.(18). }
\end{figure}

In what follows, we turn our attention to 
the contribution to the spectrum  coming from the
Hamiltonian $H^4(k)$ in the presence of the magnetic field $\mathcal{B}$.
As already mentioned, the case $B=0$ was studied perturbatively 
in \cite{Ezawa}, where one proves that $t_4\neq 0$ generates 
the dispersion of the band, which thus becomes quasi-flat.
In our calculation, we shall consider a large M and neglect the 
splitting $\delta$ (which is anyhow much smaller than the band
dispersion). Then, the eigenvalues in the presence of 
the hopping $t_4$ and of the magnetic field $\mathcal{B} \neq 0$  
will be given by  the equation: 
\begin{equation}
\begin{vmatrix}
E-<\Psi_A|H^4 |\Psi_A> &<\Psi_A|H^4 |\Psi_B>\\
<\Psi_B|H^4 |\Psi_A> & E-<\Psi_B|H^4 |\Psi_B>\\
\end{vmatrix}
=0.
\end{equation}
Since $<\Psi_A|H^4|\Psi_B>=0$ the result reads:
\begin{eqnarray}
E^1(k)=<\Psi_A(k)|H^4|\Psi_A(k)>=\sum_{m=1}^M  
\overline{t_{4A}}(m)\xi^{A*}_{k,m+1}\xi^{A}_{k,m} + c.c.
\nonumber\\
E^2(k)=<\Psi_B(k)|H^4 |\Psi_B(k)>=\sum_{m=1}^M  
\overline{t_{4B}}(m)\xi^{B*}_{k,m+1}\xi^{B}_{k,m} + c.c. ,
\end{eqnarray}
with the notation  $\overline{t}_{4A}(m)=t_4 \big(e^{i\phi_{4A}(m)}+
e^{-i(\phi_{4A}(m)-k)}\big)$, and a similar one for 
$\overline{t}_{4B}(m)$.

For zero magnetic flux, in the limit $M\rightarrow \infty$, 
Eq.(16) yields Ezawa's result: 
\begin{equation}
E^1(k)=E^2(k)= -4\frac{t_4 t_1}{t_2}(1+cos k) ,
\end{equation}
saying that the levels  remain degenerate but depend on $k$, 
such that they get a dispersion equal to $8t_4t_1/t_2$. 
However, the interesting case occurs at $\Phi \neq 0$ when the 
degeneracy is lifted. The exact summation in Eq.(16) is difficult, 
so  we approximate  it by taking advantage of the strong 
localization of the coefficients $\xi_{k1}^{A}$ and $\xi_{kM}^B$ 
at the edges $m=1$ and $m=M$, respectively. Using also  Eq.(13)
one obtains: 
\begin{eqnarray}
E^1(k,\Phi)&\cong&\overline{t}_{4A}(1,\Phi)\big(-\frac{\overline{t_1(1)}^*}
{t_2}\big)
|\xi^A_{k,1}|^2 + c.c.
\nonumber\\
E^2(k,\Phi)&\cong&\overline{t}_{4B}(M-1,\Phi)\big(-\frac{\overline{t_1(M)}}{t_2}
\big)|\xi^B_{k,M}|^2 + c.c.~ ,
\end{eqnarray}
with $E^1(k,\Phi)\neq E^2(k,\Phi)$, indicating the degeneracy lifting 
due to the magnetic field.

Figure 5 compares the quasi-flat spectrum in the absence (left panel)
and presence (right panel) of the magnetic field applied on the ribbon. 
While the hopping $t_4$ generates the dispersion, the magnetic field 
gives rise to the degeneracy lifting. The red lines represent 
the numerical result, who considers all hopping parameters $t_1,.., t_5$, 
while the black lines represent the analytical formulas Eq.(17) 
and Eq.(18) \cite{NoteEq18} calculated with $t_3=t_5=0$. 
The fit being very good, one concludes that the hopping parameters 
$t_3$ and $t_5$ have negligible influence on  the spectrum, 
at least in the energy range of the quasi-flat band.

\section{Quantum transport and superradiance in  phosphorene  mesoscopic plaquette}
In order to investigate the transport properties in strong 
perpendicular magnetic 
field, we simulate the electronic Hall device by attaching
four leads to a finite phosphorene plaquette 
(two leads for injecting/collecting the current, and two 
voltage probes), all the leads being contacted on the same zig-zag 
edge of the  plaquette. The choice of such a lead configuration
is essential, since it is the only one that can read out the 
current carried by the topological states located close to,  and 
along the zig-zag edge.
The electron transmission coefficients between different leads, 
the longitudinal, and the transverse resistance will be calculated 
as function of a gate potential $V_{gate}$ (at a given Fermi energy in the leads)
in the Landauer-B\"uttiker formalism in terms of Green functions. 

Depending on the position of the gate potential, different types of
states become active in the transport process. Fig.3 shows the 
Hofstadter-type spectrum of the phosphorene plaquette composed of two
unequal butterflies, corresponding to the conductance and 
valence band. As usual, the Landau levels  accommodate bulk states, 
while the gaps that separate consecutive LL are filled with chiral edge 
states, induced by the quantizing magnetic field, and running all 
around the plaquette perimeter.
One may observe in Fig.3 that the chirality $dE/d\Phi$ of the edge states 
is opposite in the two bands, fact that causes the different sign  
of the quantum Hall effect in the corresponding energy ranges. 
\begin{figure}[!ht]
\includegraphics[scale=0.8]{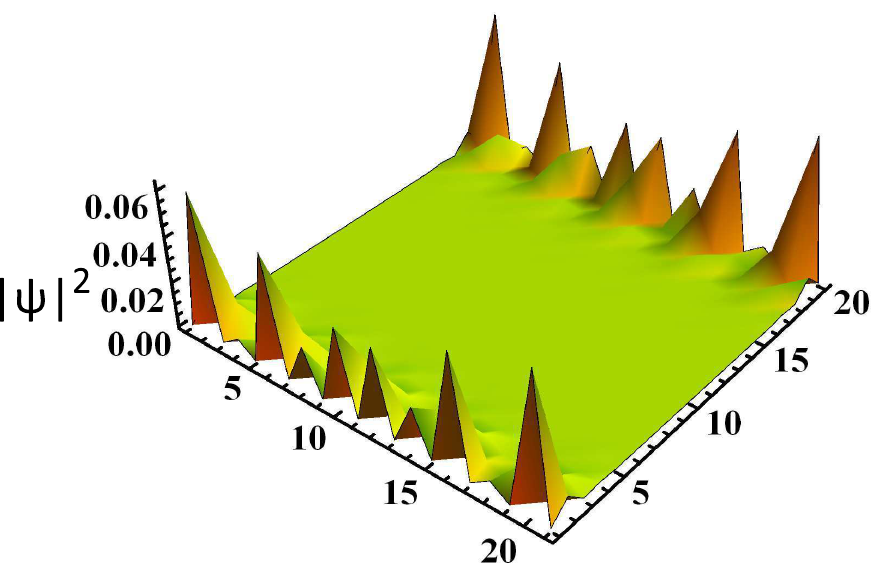}
\caption{$|\Psi|^2$  for a pair of quasi-degenerate edge states of a finite-
size plaquette in perpendicular magnetic field. 
The number of lattice sites is $21\times20$,
and the magnetic  flux is $\Phi=0.1\Phi_0$.}
\end{figure}

One has to remark the presence in the semiconducting gap of a  narrow, 
practically dispersionless band that accommodates also edge states, 
but of topological origin. They lie along the zig-zag edges only, 
exist also at $\mathcal{B}=0$, being the analogous of the edge 
states in the zig-zag ribbon discussed in the previous Section. 
It is important to underline that they do not get closed 
even  if the magnetic field is applied, 
looking  as in Fig.6.  
Figure 6 shows the superposition of two quasi-degenerate  edge states 
located near the two (left and right) zig-zag boundaries. 
Any perturbation (as a small staggering $E_A\neq E_B$, impurity disorder or 
coupling to leads) lifts  the superposition, and the wave functions 
become localized either on the left or right edge. 

Two other striking features of the 
topological edge states on the plaquette will be proved here: 
i) the dissipative character, and ii) the splitting  of the 
density of states, and the formation of minibands if the finite  
system is opened by attaching  contacts; this behavior can be 
interpreted as a {\it superradiance} effect.

\subsection{The effective Hamiltonian and transport formalism}
In order to calculate the transport quantities (namely, the longitudinal
and transverse  resistance) one needs to attach four
leads to the finite-size plaquette. Then, the Hamiltonian of the
entire system reads:
\begin{equation}
\mathcal{H}=H^S + H^L + \tau H^{LS},
\end{equation}
where the  first term is the Hamiltonian (2) of the phosphorene 
plaquette,  the second term represents all the four semi-infinite 
leads (also in the tight-binding description), and the last one  
describes the coupling between the plaquette and the leads. 
The longitudinal and Hall resistances will be calculated as function
of a gate potential $V_{gate}$ applied on the plaquette, 
similar to the experimental measurement, where
$V_{gate}$ is simulated by a diagonal term in the Hamiltonian $H^S$.

A powerful tool to deal with such an  open system is the formalism 
of the {\it effective} Hamiltonian, which is obtained by removing
the degree of freedom of the leads, with the price of losing the 
Hermicity:
\begin{equation}
H^S_{eff}=H^S+\frac{\tau^2}{t_L}~ e^{-ik}\sum_{\alpha}
|\alpha\rangle\langle\alpha|~,
\end{equation}
where $t_L$ is the hopping parameter of the tight-binding model 
for the leads, $k$ parametrizes the energy in the leads, 
$E=2t_L cos k $, and  $\{|\alpha\rangle\}$ are those localized 
states that correspond to the sites on the plaquette
where the leads are sticked to the sample \cite{Note-lead}. 
The difference between Hamiltonians Eq.(19) and Eq.(20) is just formal, 
and they are completely equivalent. The deduction of the effective 
Hamiltonian can be found for instance in Ref.6.

After constructing the matrix of the effective Hamiltonian in the 
representation of the localized functions $\{|nm\rangle\}$, 
one may calculate immediately  the  Green function  
$G(E)= (E-H_{eff}^S)^{-1}$, which enters the Landauer-B\"{u}ttiker 
formula for the {\it transmission} coefficients:
\begin{equation}
T_{\alpha\beta}=4\tau^4 |G_{\alpha\beta}|^2 Im g^L_{\alpha}Im g^L_{\beta} 
~,~~~~ \alpha\neq\beta~, 
\end{equation}
where $\alpha$ and $\beta$ ($\alpha,\beta = 1,..,4$) are lead indexes,
and Im $g^L_{\alpha}$ represents the density of states in the lead 
$\alpha$. The transmission coefficients
$T_{\alpha\beta}$ being known, the {\it conductance matrix }
$g_{\alpha\beta}$, which connects all the currents $I_{\alpha}$ 
passing through the system to the corresponding voltages $V_{\beta}$,
$I_{\alpha}=\sum_{\beta} g_{\alpha\beta} V_{\beta}$,
can be obtained as $g_{\alpha\beta}=(e^2/h) T_{\alpha\beta}$ 
for $\alpha\neq\beta$, while the diagonal conductance $g_{\alpha\alpha}$ 
results from the current conservation rule 
$\sum_{\alpha} g_{\alpha\beta}=0$.
Finally, the Landauer-B\"uttiker formalism provides the following 
expressions for the quantities of interest (longitudinal and Hall 
resistance), which are to be calculated numerically:
\begin{eqnarray}
\nonumber R_L&=&R_{14,23}=(g_{24}g_{31}-g_{21}g_{34})/|D|, \\
R_H&=&(R_{13,24}-R_{24,13})/2= 
(g_{23}g_{41}-g_{21}g_{43}-g_{32}g_{14}+g_{12}g_{34})/2|D|, 
\end{eqnarray}
where D is any $3\times3$ subdeterminant of the conductance matrix. 
Since the experimental curves show the conductance instead of the
resistance, we shall do the same, and show in Fig.8 the  the Hall
and longitudinal conductance calculated as $G_H=R_H/(R_H^2+R_L^2)$
and  $G_L=R_L/(R_H^2+R_L^2)$.
\begin{figure}
\includegraphics[scale=0.35]{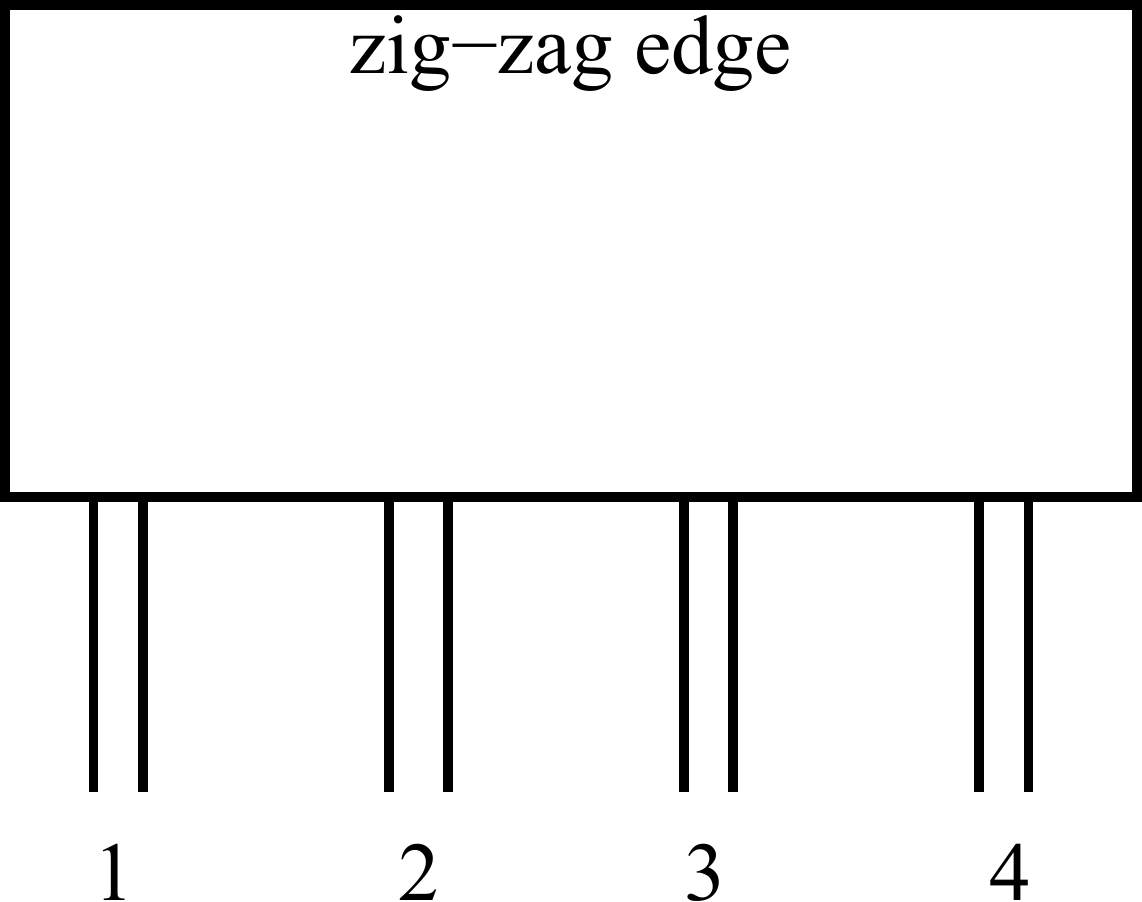}
\caption{The sketch of a four-lead Hall device. All leads are connected to the
same zig-zag edge.} 
\end{figure}

\subsection{Quantum Hall effect in phosphorene}
\begin{figure}[!ht]
\centering
\includegraphics[scale=0.8]{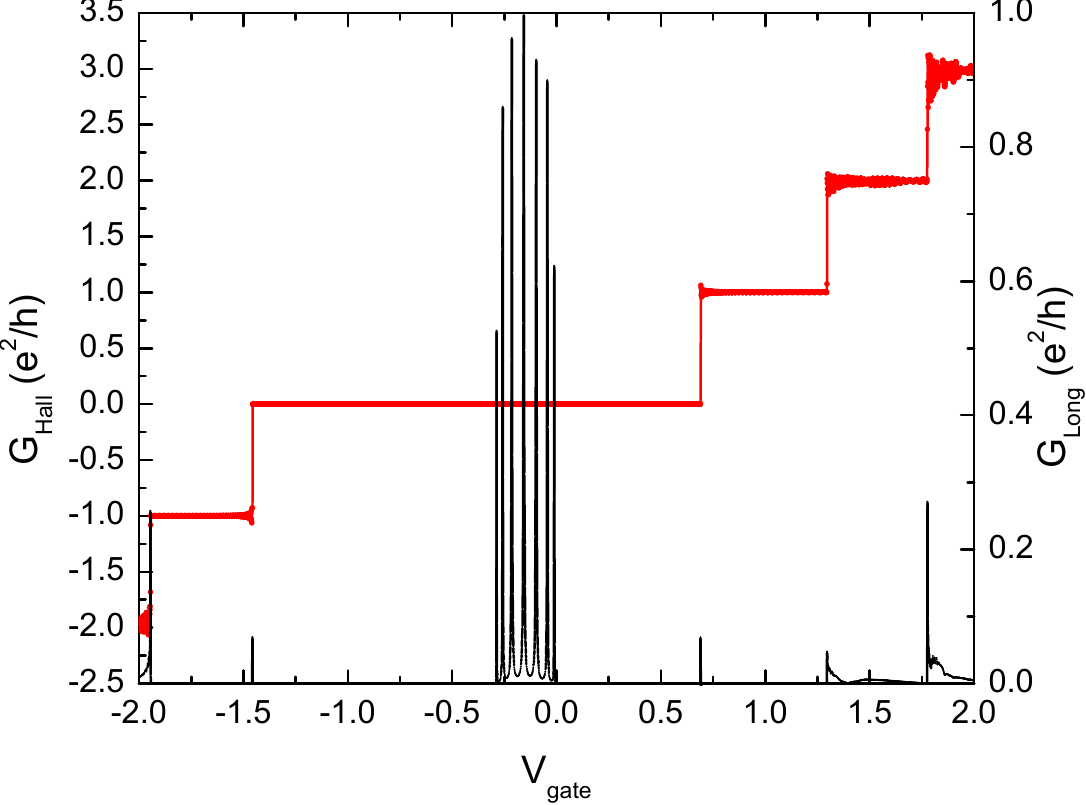}
\caption{(Color online) 
Numerically calculated  Hall (red line) and longitudinal (black line) conductance 
in the quantum Hall regime 
as a function of the gate potential $V_{gate}$, for a slightly disordered sample. 
The longitudinal conductance shows a series of peaks
in the range of the quasi-flat band.
The number of lattice sites is $107\times40$,  the magnetic flux is
$\phi/\phi_0=0.1$, and  the Fermi energy in the leads is $E_F=0$. }
\end{figure}

In what concerns the Hall conductance  in the quantum regime,  
there are significant new aspects in comparison with the graphene. 
First, one has to notice the large plateau $G_H=0$ that 
corresponds to the central gap. Next, one notice the lack of the
valley degeneracy in the low energy range,
such that the quantum Hall plateaus are the conventional (spinless) plateaus 
$n=\pm 1,\pm 2,..$ in units $e^2/h$, the same as for the two-dimensional
electron gas (2DEG) subject to a perpendicular magnetic field.
As a specific feature, one may notice in Fig.8 that the lengths 
of the plateaus in the positive and negative regions are 
slightly different, as a manifestation of the spectral asymmetry 
discussed in Sec.II.

The quantum plateaus are supported obviously by the chiral edge states
existing in the Hofstadter spectrum of the finite-size plaquette, 
however one should not forget that the central gap contains  also
topological edge states bunched in the quasi-flat band. The value $G_H=0$
everywhere in the gap confirms that these edge states are non-chiral, 
and do not support the QHE. 
Recall that, in terms of transmission coefficients, the chirality means
$T_{\alpha,\alpha+1}=integer$, while $T_{\alpha,\alpha+1}=0$ (for
any lead $\alpha$ and given direction of the magnetic field). 
On the other  hand, in the spectral range occupied by the
quasi-flat band, we find  the symmetry  
$T_{\alpha,\alpha+1}=T_{\alpha,\alpha+1}$,
which denotes the lack of chirality. This property was observed
numerically using Eq.(21), and occurs no matter the
presence or absence of the magnetic field.

The longitudinal conductance $G_L$ exhibits the 
non-dissipative behavior in the range of the quantum plateaus, 
as it should, but striking non-trivial properties are 
proved in the range [-0.3,0] covered by the quasi-flat band, 
where the longitudinal conductance is non-vanishing and shows a sequence 
of peaks.  While the dissipative character of the non-chiral edges states 
was also met in the context of the zero-energy Landau level in 
graphene \cite{Levitov1}, we think that the peaked structure of the 
longitudinal conductance is  specific to phosphorene. 

It is already known that that the flat bands are sensitive to disorder
due to their degeneracy \cite{Nita,Leykam}, 
and  one may expect that the 
$G_L$ peaks 
are also  affected by   the disorder existing in the system,
which is unavoidable experimentally. Indeed, the unitary limit $G_L=1e^2/h$
is reached only in the clean systems (this case is shown in Fig.9a), 
but any small amount of
disorder allows for the backscattering and slightly lowers the values of the 
peaks below the unitary limit. This is the case in Fig.8, where small 
Anderson (diagonal) disorder was introduced in the numerical calculation. 

In what follows we shall pay closer attention to
properties of the non-chiral edge states in the quasi-flat band.

\subsection{Spectral and transport properties of the quasi-flat band
in open system}
It is obvious that the second term of the effective Hamiltonian Eq.(20)
produces shifts of the real eigenvalues of $H^S$, but adds also an 
imaginary part, meaning the level broadening  due to the coupling 
to the leads. As long as the coupling $\tau$ is very small 
(i.e., we are in the resonant tunneling regime, which was studied for
the nanoribbon system in \cite{Schulz}), 
all the eigenvalues of $H^S$ should be practically recovered.  
However, with increasing coupling, the level broadening $\Gamma$ 
increases too, and the merging of neighboring levels occurs.
Consequently, the shape of the density of states changes significantly.
When $\Gamma \sim \Delta$ ($\Delta$ =  mean interlevel distance) 
one enters the  regime known as {\it superradiative} \cite{Zelevinsky}. 
As we already mentioned the superradiance phenomenon consists in the 
overlapping and segregation of eigenenergies  occurring in  open systems 
under the control of the  coupling between the finite system and  
the infinite reservoir.
One may expect that the energy spectrum is not everywhere equally 
sensitive to this effect, and one may  speculate that the energy 
range occupied by the states located near edges 
(where the leads are attached) is mostly affected. 

We assume that the superradiance is the mechanism that gives rise 
to the miniband structure of the quasi-flat band shown in Fig.9a, 
where the density of states ($DOS=-\frac{1}{\pi}Tr G$) exhibits eleven peaks. 
The  confirmation comes from the  calculation of the complex 
eigenvalues of the effective Hamiltonian (20). In Fig.9b we show the real 
and imaginary part of the eigenvalues in the energy range of the 
quasi-flat band, and find the presence of eleven energies with 
large imaginary part, which perfectly correspond to the positions 
of the minibands in the density of states.
One has to observe in Fig.9b also the multitude of eigenstates
with vanishing imaginary part ($ImE=0$). They correspond to the
edge states localized along the edge {\it opposite} to that one where 
the leads are connected. In other words, the process of overlapping 
and segregation affects only those edge states that are in the 
immediate vicinity of  the leads, the other ones remaining unchanged. 
\begin{figure}[!ht]
\centering
\includegraphics[scale=0.7]{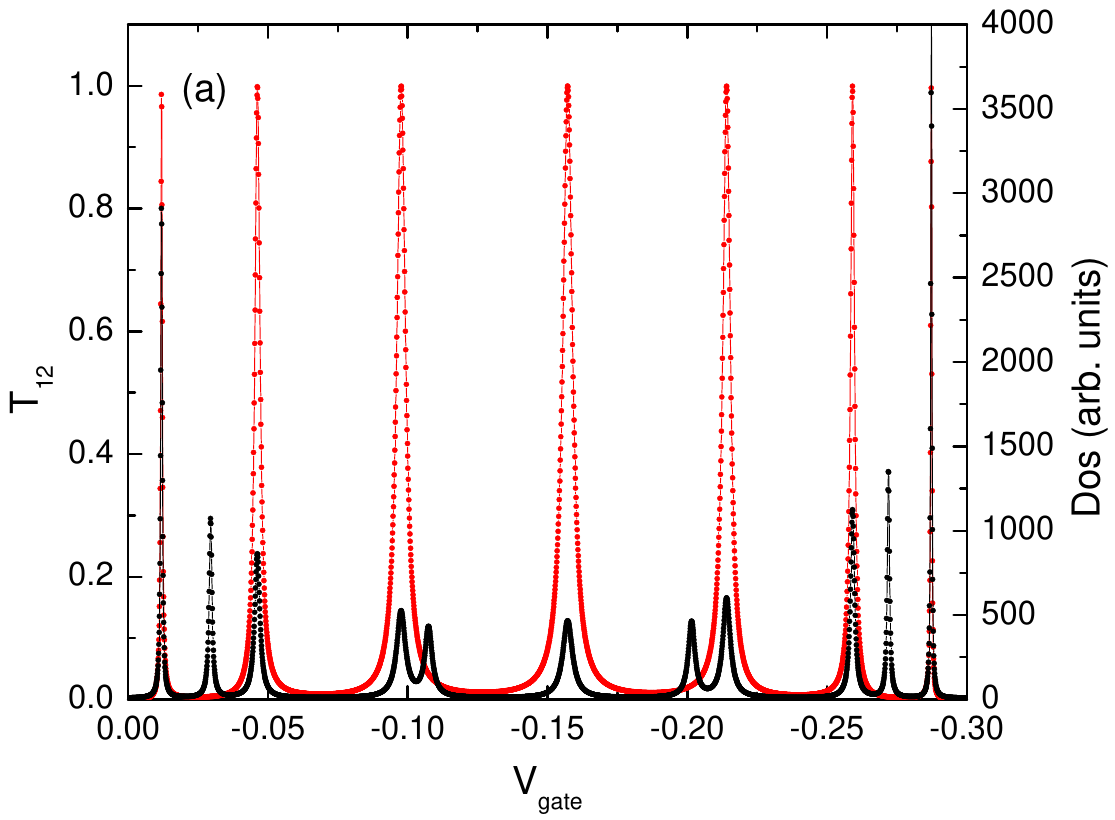}
\vskip-0.01cm
\hskip-1.20cm
\includegraphics[scale=0.7]{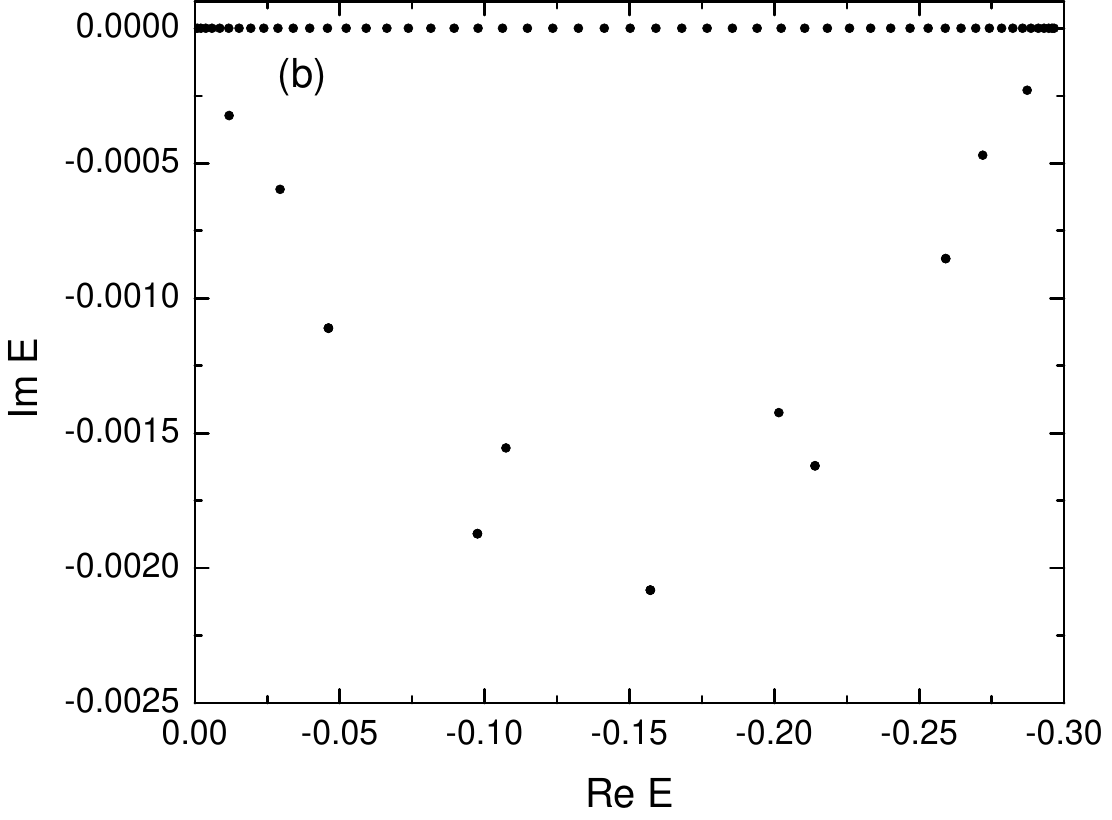}
\caption{(Color online) (a)
The peaked structure of the transmission coefficient  $T_{12}$ (red line)
and  the density of states (black line)  
of  the phosphorene plaquette in the energy range of the quasi-flat band 
for a clean system. Note that not all the DOS peaks are transmitting, 
and    also that
the unitary limit of the transmission is reached. 
(b) $Im E$ vs $Re E$ for the eigenvalues of the effective Hamiltonian (20) 
corresponding to the quasi-flat band. Note that  the eigenvalues with 
$Im E \neq 0$ correspond to the miniband structure in the panel (a).
The number of lattice sites is $107\times40$,  the magnetic flux is
$\Phi=0.1\Phi_0$,  the Fermi energy in the leads is $E_F=0$, and $\tau=2 eV$.}
\end{figure}

As a next step, we draw the attention to the fact, visible in Fig.9a, 
that {\it not all}  DOS peaks support the electron transmission. 
This apparently surprising result has a simple explanation in terms of the 
charge distribution on the plaquette, described by the 
local density of states. The local density of states, calculated at 
each site $i$ on the plaquette as the imaginary part of the Green function
$LDOS_i(E)=-\frac{1}{\pi} Im G_{ii}(E)$,
shows that, in the case of the conducting minibands, the states are located 
between  the contacts, but, for the non-conducting ones, the states 
are positioned outside the contacts. The two situations are displayed 
in Fig.10, The same figure tells furthermore
that the  states accommodated by the
minibands do not get closed around the whole perimeter of the
plaquette, even at such  strong magnetic fields that generates 
chiral edge states in the bands.

\begin{figure}[!ht]
\centering
\includegraphics[scale=0.7]{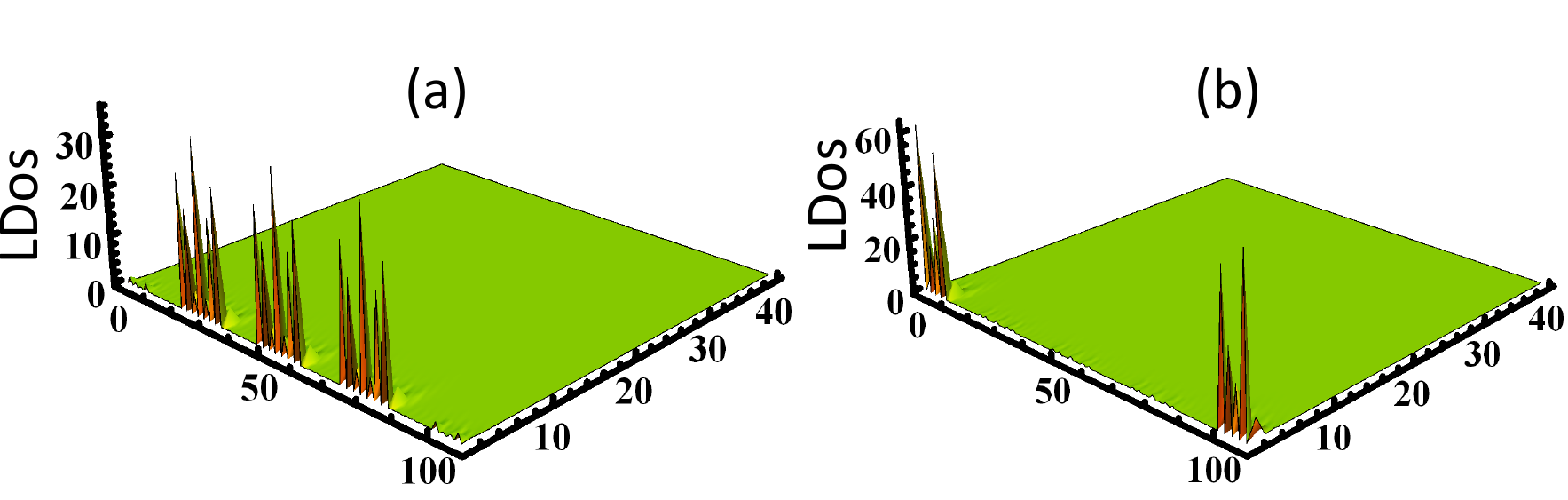}
\caption{(Color online): The local density of
states (LDOS) in the presence of the leads: 
(a) the density of states is localized in-between the leads 
and contribute to transmission (corresponding to the peak at
$V_{gate}=-0.0978 eV$ in Fig.9a),  and (b) the density of states is localized
outside the leads and corresponds to the non-transmitting DOS peak at
$V_{gate}=-0.10735 eV$ in Fig.9a.
The parameters are the same as in Fig.9.}
\end{figure}

In what concerns the transport properties,
besides the lack of chirality mentioned above, we find that
the transmission exhibits a peaked structure and reaches the 
unitary limit $T_{\alpha\alpha+1}=T_{\alpha+1\alpha}=1 $ 
in the middle of the conducting minibands. 
This  behavior of the transmission coefficients is proved by numerical 
investigation  using Eq.(21), and it is shown in Fig.9a.
(Of course, the unitary limit, telling that each miniband behaves 
as a perfect one-dimensional channel, is reached only for 
disorder-free systems.)
All the other transmission coefficients vanish, so that
the whole $4 \times 4$ conductance matrix reads as:
\begin{equation}
\text{\textsl{\bf g}} =
\frac{e^2}{h}
\begin{pmatrix}
  -1 & ~1 & 0 & 0 \\
   ~1 & -2 & 1 & 0 \\
   0 & 1 & -2 & 1 \\
   0 & 0 & 1 & -1 
\end{pmatrix},
\end{equation}
and allows for the calculation of the Hall ($R_H$) and 
longitudinal ($R_L$)  resistances.
Indeed,  by the use of Eq.(22), one obtains the results already known from the
numerical calculation. 
For the Hall resistance one gets  $R_H=0$, which is the outcome of the
lack of chirality,
however, a non-trivial result is obtained for the
longitudinal response, for which the above matrix yields $R_L=1 h/e^2$, 
indicating a dissipative character of the electron transport in minibands.
It is to  underline that this distinctive property of the quasi-flat band 
occurs even in the presence of a strong magnetic field, which, otherwise, is 
able to  generate in the other bands the specific QHE behavior, 
i.e., quantized non-zero values of $R_H$, and non-dissipative $R_L$.

\section{summary  and conclusions}
In this paper we have studied spectral and transport properties of 
phosphorene, paying special attention to confined systems 
(zig-zag ribbon and mesoscopic plaquette) subject to a magnetic field, 
with main focus on the topological edge states organized in the  quasi-flat 
band. Our results are the following:

 We approach analytically the question of electron-hole symmetry breaking,
and demonstrate the role played in this respect by  the hopping integral $t_4$,
the only parameter in the tight-binding model that violates
the bipartitism of the lattice.

 The Hofstadter-type spectrum of the phosphorene plaquette misses the
usual periodicity $E(\Phi)=E(\Phi+\Phi_0)$ because  three different  
Peierls phases (depending also on the quasi-2D lattice angle $\beta$) 
are assigned to different hopping terms.
The Hofstadter spectrum comprises edge states of chiral and topological origin. 
The chiral states fill the gaps between the Landau levels
and extend all around the perimeter. The other ones extend along the 
zig-zag edges only, and remain like that even in strong  magnetic field. 
The topological states are bunched in a quasi-flat band located in the 
middle of the gap. 

 For the zig-zag ribbon, since  $|t_1/t_2|<1$, we prove that 
the topological edge states occur at any momentum $k$ in the 
Brillouin zone,  contrary to the graphene case.  
We  analytically show that the degeneracy of a pair of 
edge states, located at opposite edges of the ribbon, occurs only in 
the limit of the infinite wide ribbon ($M\rightarrow\infty$), 
and we prove also that the degeneracy is lifted by the magnetic field.

 The  quantum transport in the mesoscopic plaquette 
is treated numerically. The Hall device may use different lead configuration,
however the configuration 'all leads on the same edge' (Fig.7) is that one that
evidence better the features of the edge states. We suggest such a 
configuration for an eventual  experimental study of the topological edge states.
Specific to phosphorene, the Hall conductance shows a zero 
plateau in the gap, indicating the non-chiral behavior of  the quasi-flat band, 
but a non-zero longitudinal  conductance,  indicating the dissipative character.
The multiple-peak aspect of $G_L$ reflects the miniband structure of the 
density of states in the presence of the leads. 

 For the energy range occupied by the quasi-flat band, we calculate 
the transmission coefficient between consecutive leads, 
the DOS of the plaquette when connected to the leads, and the complex eigenenergies
of the non-Hermitian effective Hamiltonian. The ensemble of these quantities
(shown in Fig.9),
which  are controlled by the dot-lead coupling parameter $\tau$, 
certifies the  manifestation of the superradiance phenomenon 
in phosphorene. 
We underline that not all the minibands in the DOS are conducting (Fig.9a),
the issue being explained by Fig.10.
We mention the features of the electron transmission, which 
besides the lack of chirality, exhibits unitary 
peaks in the case of the clean system, proving a  one-channel- type transport.

 In what concerns the disorder effects, it is clear that the different
quantum states respond differently to disorder. While the chiral states are
robust, one expects the topological edge states be sensitive due to their 
quasi-degeneracy. We think that the localization, level spacing analysis, and
the electron transmission are topics of interest, which are however 
beyond the aim of this paper. 

In conclusion,  phosphorene is a 'beyond' graphene material which, besides
potential applications as a semiconductor,  shows  several
interesting conceptual  properties, mainly  concerning the topological 
edge states accommodated in the quasi-flat band.

\section{Acknowledgments}
We are grateful to Antonio Castro Neto for  illuminating discussions on the 
phosphorene issue, and to Marian Nita for helpful comments.
We acknowledge financial support from PNII-ID-PCE Research Programme
(grant no 0091/2011) and Romanian Core Research Programme.


\begin{thebibliography}{30}
\bibitem{Note-phase} The  particular expression of the
phases depend on the choice of the origin; in our case they 
correspond to the choice in Fig.1
\bibitem{NoteW} We remind that in the graphene ribbon, since $t_1=t_2$, the convergence, 
and, consequently, the flat band occur only for $k$ comprised between the
points $K$ and $K'$ in the Brillouin zone \cite{Wakabayashi} .
\bibitem{NoteEq18} In Fig.5(right), the black lines represent Eq.(18), where 
$|\xi_{k1}^A|=|\xi_{kM}^B|=1$ has been considered.
\bibitem{Note-lead} One has to keep in mind the technical detail
that each lead $\alpha=1,.,4$
is composed of several one-dimensional independent channels, 
and that in Eq.(20) the summation over  all the channels is implicit. 
The same convention is assumed also in Eq.(21).
\end{thebibliography}
\end{document}